\documentclass[aps,prb,twocolumn,showpacs]{revtex4}
\usepackage[dvips]{graphicx}

\begin{document}

\title{Fermi surface evolution in the antiferromagnetic 
state for the electron-doped $t$-$t'$-$t''$-$J$ model}
\author{Qingshan Yuan,$^{1,2}$ Yan Chen,$^3$ T. K. Lee,$^4$ and C. S. Ting$^1$}
\address{$^1$Texas Center for Superconductivity and Department of Physics,
University of Houston, Houston, TX 77204\\
$^2$Pohl Institute of Solid State Physics, Tongji University,
Shanghai 200092, P. R. China\\
$^3$Department of Physics, University of Hong Kong, Pokfulam Road, Hong Kong,
P. R. China\\
$^4$Institute of Physics, Academia Sinica, Nankang, Taipei, Taiwan 11529
}

\begin{abstract}
By use of the slave-boson mean-field approach, we have studied the electron-doped
$t$-$t'$-$t''$-$J$ model in the antiferromagnetic (AF) state.
It is found that at low doping the Fermi surface (FS) pockets appear around 
$(\pm\pi,0)$ and $(0,\pm\pi)$, and upon increasing doping the other ones will 
form around $(\pm{\pi\over 2},\pm{\pi\over 2})$. The evolution of the FS 
with doping as well as the calculated spectral weight are consistent with 
the experimental results. 

\end{abstract}

\maketitle

Recent angle-resolved photoemission spectroscopy (ARPES) measurements
\cite{Armitage02,Damascelli} have revealed the doping evolution of
the Fermi surface (FS) in electron-doped cuprate
Nd$_{2-x}$Ce$_x$CuO$_4$. It was found that at low doping a small FS pocket
appears around $(\pi,0)$, in contrast to the hole-doped case where the low-lying
states are centered at momentum $({\pi\over 2},{\pi\over 2})$. 
Upon increasing doping another pocket begins to form around 
$({\pi\over 2},{\pi\over 2})$ and eventually at optimal doping $x=0.15$ 
the several FS pieces evolve into a large curve around $(\pi,\pi)$.

The ARPES data provide clear evidence for electron-hole asymmetry in high-temperature
superconductors. This asymmetry has already been seen in the temperature/doping phase 
diagram, where the antiferromagnetic (AF) phase is much more robust and 
at the same time the superconducting (SC) one is much narrower in the electron-doped 
materials. To understand it, the role of the long-range hoppings $t'$ and $t''$ 
was stressed.
Based on the $t$-$t'$(-$t''$)-$J$ model, the single-hole/electron
behavior and its excitation spectrum have been studied by various methods.
\cite{Tohyama94,Gooding,Xiang,Lee97,Tohyama01}
It was shown that a doped hole enters into the lowest
energy state at momentum $({\pi\over 2},{\pi\over 2})$, while a doped electron 
goes into that at $(\pi,0)$. The excitation spectrum for slightly hole/electron-doped 
$t$-$t'$-$t''$-$J$ model has also been investigated by variational Monte Carlo 
technique,\cite{Lee03} which provides an explanation for the different FS structure in 
the hole- and electron-doped cuprates at low doping. In all these studies, the 
electron-doped Hamiltonian has been mapped onto the hole-doped one with extra 
minus signs for the hopping parameters,\cite{Tohyama94,Lee03}
thus both electron- and hole-doped systems have been treated in the same manner.

On the other hand, it was claimed recently by Kusko {\it et al.}\cite{Kusko} 
that the use of $t$-$U$ type models is essential in the study of 
electron-doped cuprates. 
By mean-field (MF) treatment on the $t$-$t'$-$t''$-$U$ model they have found 
that the FS evolution with doping in the AF state is consistent with the ARPES 
results. But in their theory the on-site $U$ is treated as a 
doping-dependent effective parameter phenomenologically.

It is not clear whether a different Hamiltonian should be constructed when 
electron-doped cuprates are studied. At least one would ask if the 
$t$-$t'$-$t''$-$J$ model could have reproduced the FS evolution in the 
electron-doped case (even without argument of any doping-dependent parameter as 
done by Kusko {\it et al.}).
As already stated above, for single or slightly electron-doped case the 
$t$-$t'$-$t''$-$J$ model has given good results compared with the experiments.
Naturally the same model should be further studied at {\it finite} doping.

Another concern for the choice of $t$-$J$ type models is the SC
pairing symmetry. It is now generally believed that the SC gap in hole-doped cuprates
has a $d$-wave symmetry. And the theoretical studies were substantially based on the
$t$-$J$ type models. Although not well clarified, it has been strongly suggested
by various experimental measurements such as 
phase sensitive,\cite{Tsuei} tunneling,\cite{Biswas} ARPES,\cite{Armitage01} 
penetration depth,\cite{Kokales,Prozorov,Skinta} etc. 
that the pairing symmetry is also $d$ wave for electron-doped cuprates at low and 
optimal doping. Thus one would expect that
the $d$-wave superconductivity in both electron- and hole-doped cuprates may be
understood by a {\it unified} $t$-$J$ type model.

In this paper we study the $t$-$t'$-$t''$-$J$ model by use of the slave-boson 
MF approach in the electron-doped case. In consideration of the AF order, 
the energy bands and the corresponding FS are calculated
in the magnetic Brillouin zone (MBZ).
The experimentally observed FS evolution with doping is reproduced, essentially,
a pocket appears first around $(\pi,0)$, and another one shows up around
$({\pi\over 2},{\pi\over 2})$ upon increasing doping.

The $t$-$t'$-$t''$-$J$ model Hamiltonian is written as
\begin{widetext}
\begin{eqnarray}
H & = & -t\sum_{\langle ij\rangle \sigma}(c_{i\sigma}^{\dagger}c_{j\sigma}+{\rm h.c.})
-t'\sum_{\langle ij\rangle_2\sigma}(c_{i\sigma}^{\dagger}c_{j\sigma}+{\rm h.c.})
-t''\sum_{\langle ij\rangle_3 \sigma}
(c_{i\sigma}^{\dagger}c_{j\sigma}+{\rm h.c.})\nonumber\\
& & +J\sum_{\langle  ij\rangle }(\vec{S}_i \cdot \vec{S}_j-{1\over 4}n_i n_j)
-\mu_0 \sum_{i\sigma}c_{i\sigma}^{\dagger}c_{i\sigma}\ ,\label{H}
\end{eqnarray}
\end{widetext}
where 
$\langle\rangle,\ \langle\rangle_2,\ \langle\rangle_3$
represent the nearest neighbor (n.n.), second n.n., and third n.n. sites, 
respectively, and the rest of the notation is standard. 
No double occupancy is allowed in the model. The Hamiltonian is essentially
the same as that in the hole-doped case except here for electron-doping, 
$t<0,\ t'>0$ and $t''<0$.\cite{Tohyama94,Gooding,Tohyama01}
Throughout the work $|t|$ is taken as the energy unit.
Since we mainly focus on the qualitative results, no model parameters
are fine-tuned. Typical values are adopted: $t'=0.3,\ t''=-0.2$ and $J=0.3$.

We make use of the slave-boson transformation, i.e.,
$c_{i\sigma} =b_i^{\dagger} f_{i\sigma}$
with $b_i$: bosonic holon operator, $f_{i\sigma}$: fermionic spinon operator
and the constraint $b_i^{\dagger}b_i+\sum_{\sigma}f_{i\sigma}^{\dagger}f_{i\sigma}=1$
at each site. Since we are interested in low temperatures,
boson condensation is assumed, 
i.e., $\langle  b_i\rangle =
\langle  b_i^{\dagger}\rangle =\sqrt{\delta}$ ($\delta$: doping concentration).
Then we decouple the Hamiltonian by defining the MF parameters:
$$\langle  f_{i\sigma}^{\dagger}f_{i\sigma}\rangle =(1-\delta)/2+\sigma(-1)^i m$$
[thus $m=(-1)^i\langle S_i^z\rangle$ representing the AF order]
and the uniform bond order
$$\langle  f_{i\sigma}^{\dagger}f_{j\sigma}\rangle =\chi\ .$$
Based on the above treatment, the Hamiltonian (\ref{H}) can be 
expanded as follows in momentum space (up to irrelevant constants)
\begin{eqnarray}
H & = & \sum_{k,\sigma} (\varepsilon_k f_{k\sigma}^{\dagger}f_{k\sigma}
+ \varepsilon_{k+Q} f_{k+Q\sigma}^{\dagger}f_{k+Q\sigma})\nonumber\\
& & -2Jm \sum_{k,\sigma} \sigma (f_{k\sigma}^{\dagger}f_{k+Q\sigma}+{\rm h.c.})
+2NJ(\chi^2+m^2)\ ,\nonumber\\ \label{Hk}
\end{eqnarray}
where $\varepsilon_k=(2|t|\delta-J\chi) (\cos k_x+\cos k_y)
-4t'\delta\cos k_x \cos k_y-2t''\delta(\cos 2k_x+\cos 2k_y)-\mu$, 
$Q=(\pi,\pi)$, and $N$ is the total number of lattice sites.
The chemical potential is renormalized: $\mu=\mu_0-\lambda+J(1-\delta)$
with $\lambda$: the Lagrange multiplier. 
(The local constraint has been relaxed to the global one.)
Note that the summation to $k$ is over the MBZ:
$-\pi<k_x\pm k_y\le \pi$.

By use of the following unitary transformations:
\begin{equation}
\left(\begin{array}{l} f_{k\sigma}\\ f_{k+Q\sigma}\end{array}\right)
=\left( \begin{array}{cc}
\cos \theta_k & \sigma\sin \theta_k\\
\bar{\sigma}\sin \theta_k & \cos \theta_k \end{array}
\right)
\left(\begin{array}{l} \alpha_{k\sigma}\\ \beta_{k\sigma}\end{array}\right)\ ,
\end{equation}
with 
\begin{equation}
\cos 2\theta_k = (\varepsilon_{k+Q}-\varepsilon_{k})/
\sqrt{(\varepsilon_{k+Q}-\varepsilon_{k})^2+4(2Jm)^2}\ ,
\end{equation}
the Hamiltonian (\ref{Hk}) can be diagonalized into
\begin{eqnarray}
H & = & \sum_{k\sigma} (\xi_{k\alpha}\alpha_{k\sigma}^{\dagger}\alpha_{k\sigma}
+\xi_{k\beta}\beta_{k\sigma}^{\dagger}\beta_{k\sigma}) + 2NJ(\chi^2+m^2)\ ,
\nonumber\\
\end{eqnarray}
with the energy bands
\begin{eqnarray}
\xi_{k\alpha,\beta} & = & (\varepsilon_{k}+\varepsilon_{k+Q})/2
\mp \sqrt{(\varepsilon_{k+Q}-\varepsilon_{k})^2/4+(2Jm)^2}\ .
\nonumber\\
\end{eqnarray}
The free energy is simply given by ($k_B=1$)
\begin{eqnarray}
F & = & -2T \sum_{\eta=\alpha,\beta} \sum_{k}
\ln (1+e^{-\xi_{k\eta}/T})+ 2NJ(\chi^2+m^2)\ .
\nonumber\\
\end{eqnarray}
The MF parameters $m$ and $\chi$ are then decided by
$\left({\partial F\over \partial m}\right)_{\mu}=0$ and 
$\left({\partial F\over \partial \chi}\right)_{\mu}=0$. 
The chemical potential $\mu$ is adjusted to yield the right filling:
$-{\partial F\over \partial \mu}=N(1-\delta)$. 
For each given doping $\delta$ and temperature $T$, the quantities
$m$, $\chi$ and $\mu$ are calculated self-consistently.

\begin{figure}[ht]
\begin{center}
\includegraphics[width=7cm,height=5cm,clip]{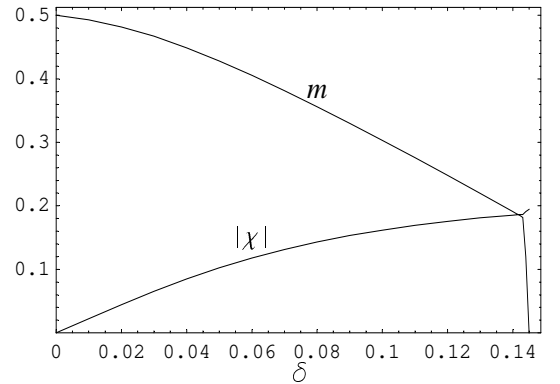}
\end{center}
\caption{The MF parameters $m$ and $\chi$ (negative) as functions of $\delta$
at $T=0.001|t|$. The model parameters are taken as: 
$t'=0.3$, $t''=-0.2$, $J=0.3$ (in units of $|t|$).}
\label{Fig:mchi}
\end{figure}

In Fig.~\ref{Fig:mchi} we give the results for $m$ and $\chi$ as functions of
doping at $T\simeq 0$. It is seen that the staggered magnetization decreases with 
increasing doping, and goes sharply to zero at around $\delta=0.145$. 
The AF order is overestimated due to the MF treatment: 
for example, in the undoped case ($\delta=0$), one has $m=0.5$,
which is larger than the accurate value $0.3$ for the two-dimensional 
Heisenberg model. In spite of this,
the obtained energy bands are still instructive to understand the experimental 
observation as shown below.

\begin{figure*}[ht]
\begin{center}
\includegraphics[width=12cm,height=8cm,clip]{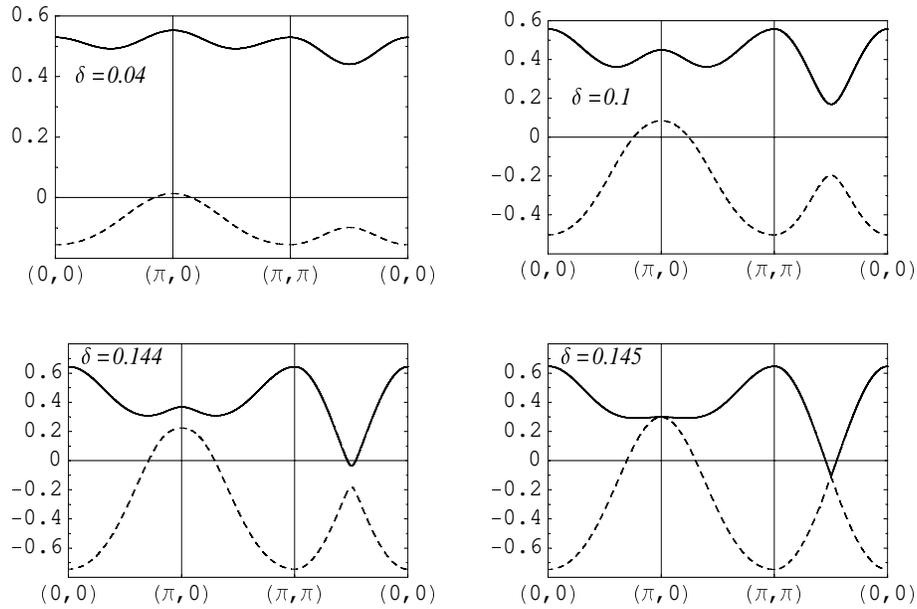}
\end{center}
\caption{The energy dispersions for different dopings. In each panel, the 
solid and dashed lines are for $\beta$ and $\alpha$ bands, respectively,
and the Fermi energy is fixed at zero.}
\label{Fig:AF}
\end{figure*}

\begin{figure*}[ht]
\begin{center}
\includegraphics[width=10.5cm,height=10cm,clip]{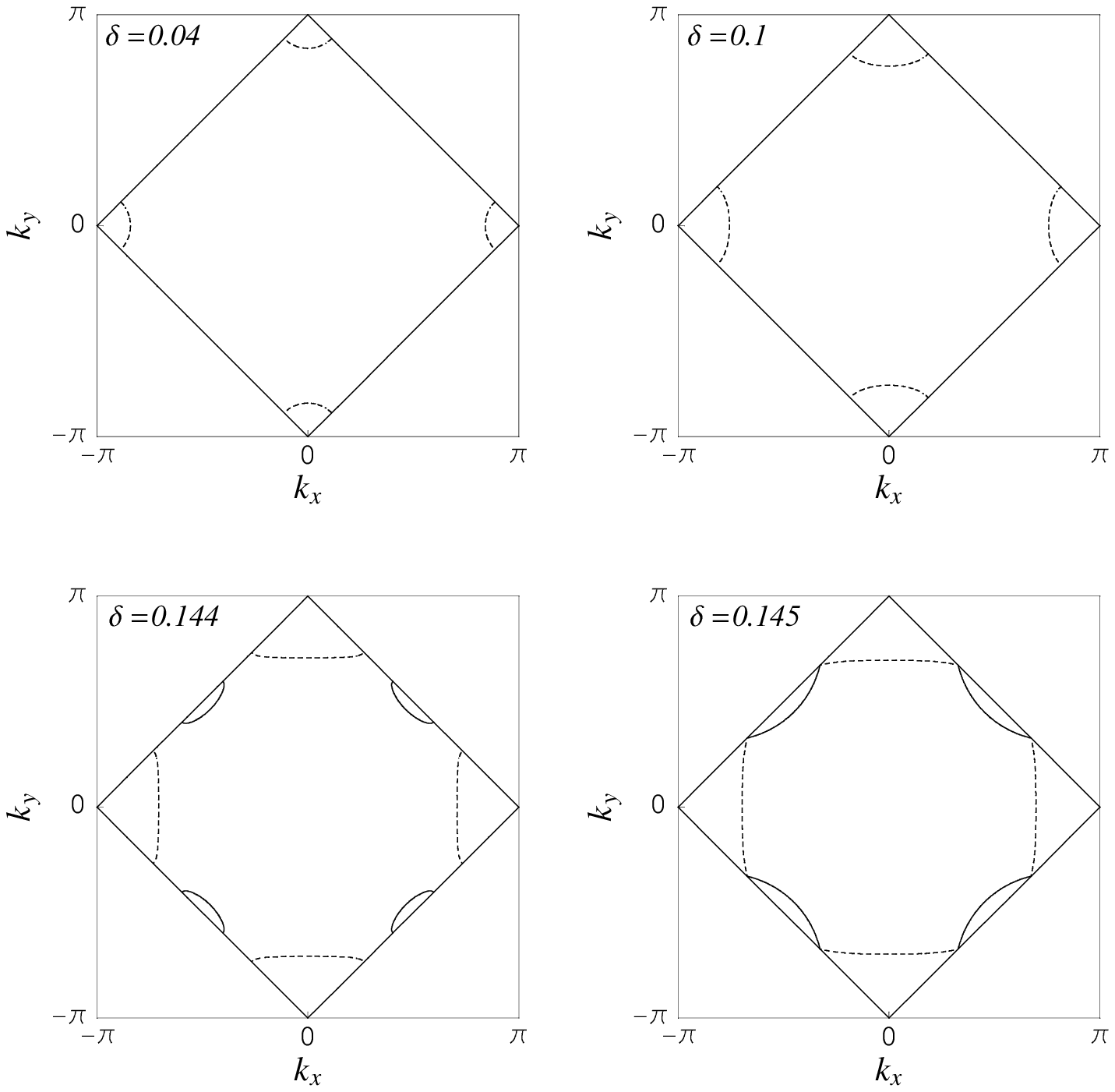}
\end{center}
\caption{The Fermi surfaces plotted in the MBZ: one-to-one
correspondence to the energy bands shown in Fig.~\ref{Fig:AF}.}
\label{Fig:FS}
\end{figure*}

\begin{figure*}[ht]
\begin{center}
\includegraphics[width=12cm,height=4cm,clip]{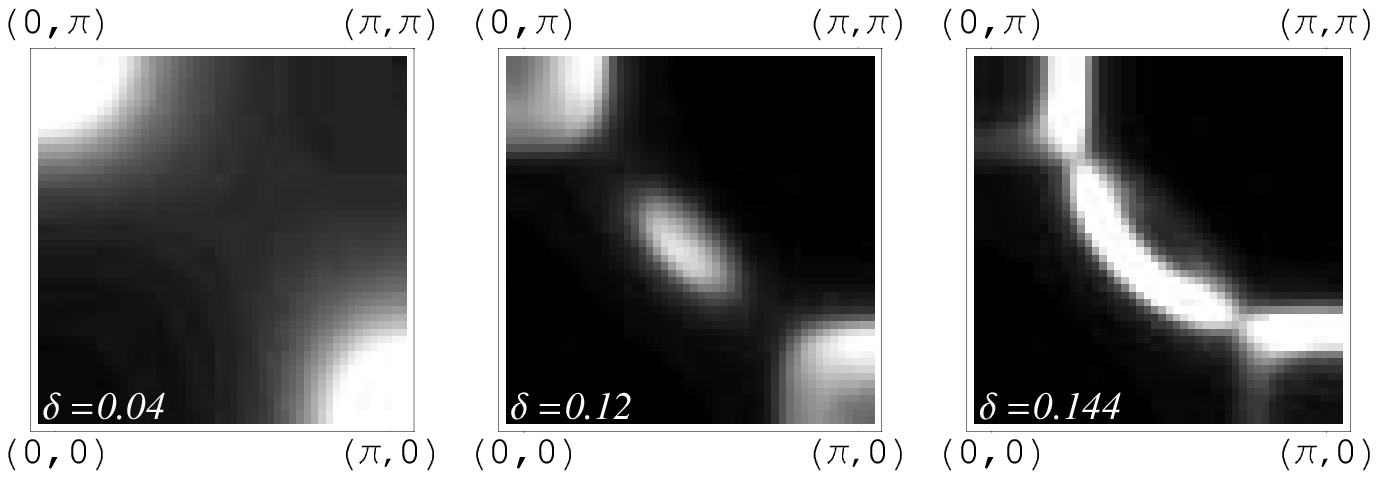}
\end{center}
\caption{The maps for spectral weight, obtained by integrating $A(k,\omega)$
times the Fermi function over an energy interval [$-0.06$, $0.12$]$|t|$ 
($\sim$ [$-20$, $40$] meV corresponding to that adopted in the ARPES experiments) 
around the Fermi level. Highs are denoted by white and lows by black.}
\label{Fig:AKW}
\end{figure*}

The energy dispersions are plotted in Fig.~\ref{Fig:AF},
and the corresponding Fermi surfaces in Fig.~\ref{Fig:FS}.
Note that the beginning Hamiltonian (\ref{H}) with $t<0$ (and $t'>0$,\ $t''<0$)
is already the transformed version after particle-hole transformation,
thus doping electron actually means doping hole in our treatment.
For the same reason, $\xi_{k\alpha,\beta}$ in Fig.~\ref{Fig:AF} should be inverted,
i.e., $\xi_{k\alpha,\beta}\rightarrow -\xi_{k\alpha,\beta}$ if
they are understood as the energy bands for the original electrons. 
The Matsubara Green's function for $f$-operator is simply written down:
$G(k,i\omega_n)=\cos^2 \theta_k/(i\omega_n-\xi_{k\alpha})+
\sin^2 \theta_k/(i\omega_n-\xi_{k\beta})$. And the single particle ($c$-electron)
spectral function is $A(k,\omega)=
-(\delta/\pi)$Im$G(k,i\omega_n)|_{i\omega_n\rightarrow \omega+i0^+}$.
The density plots for $\int {\rm d}\omega A(k,\omega)/(1+e^{-\omega/T})$ over
an energy interval around the Fermi level have been shown 
in Fig.~\ref{Fig:AKW} for a few dopings.
We come to see the details in the following.

At low doping ($\delta=0.04$), the Fermi level crosses only the $\alpha$-band.
The low-energy states are around $(\pi,0)$ and equivalent points, in consistence 
with the numerical results based
on the single/slightly electron-doped systems.\cite{Tohyama94,Gooding,Lee03}
Correspondingly small FS pockets appear around $(\pm\pi,0)$ and  
$(0,\pm\pi)$, as shown by the upper-left panel of Fig.~\ref{Fig:FS} 
(see also the left panel of Fig.~\ref{Fig:AKW}).
The energy gap above the Fermi level at $({\pi\over 2},{\pi\over 2})$
is about $0.44|t|\simeq 0.15$ eV (if typical $|t|=0.35$ eV is taken),
which is in the same energy scale $\sim 0.2$ eV as measured by 
ARPES.\cite{Armitage02,Damascelli}
With increasing doping ($\delta=0.1$), the staggered magnetization is reduced.
Then the two bands become close to each other
and the energy minimum of the $\beta$-band
at $({\pi\over 2},{\pi\over 2})$ shifts towards the Fermi level. 
The FS is still contributed by the single $\alpha$-band: the pockets
around $(\pm\pi,0)$ and $(0,\pm\pi)$ expand.
Although the $\beta$-band has not touched the Fermi level,
it may contribute enough strong spectral intensity around
$({\pi\over 2},{\pi\over 2})$ if it further approaches the latter
with increasing doping. This is evidenced by the middle panel of Fig.~\ref{Fig:AKW}
for $\delta=0.12$, where finite spectral intensity is clear around
$({\pi\over 2},{\pi\over 2})$. Also there, it is observed that
half of the square around $(\pi,0)$ loses much of its intensity,
consistent with the ARPES findings. When doping continues to increase,
e.g., for $\delta=0.144$, the two bands are strongly overlapped and both 
crossed by the Fermi level. New FS pockets centered at 
$(\pm{\pi\over 2},\pm{\pi\over 2})$ contributed by the $\beta$-band will emerge, 
as shown by the solid lines in the lower-left panel of Fig.~\ref{Fig:FS}.
If the FS curves for $\alpha$-band (the dashed lines) are moved outside of the 
MBZ by the wavevector $Q$, one will see totally three FS pieces 
in the first quadrant, which, when looked together, are close to a large curve 
around the point $(\pi,\pi)$. This is more explicitly seen from the spectral
intensity as shown in the right panel of Fig.~\ref{Fig:AKW}.
From the above, we can see that the FS evolution
with doping is in essential agreement with the ARPES measurements.
\cite{Armitage02,Damascelli}

Eventually, when $\delta=0.145$, the AF order breaks down. The two bands touch,
which will merge into a single band when plotted in the original Brillouin zone.
This is suggested by the lower-right panel of Fig.~\ref{Fig:FS},
where a single continuous FS curve will form in the first quadrant
if the dashed lines are moved outside of the MBZ by $Q$.

The energy bands at finite doping as shown in Fig.~\ref{Fig:AF} are, if inverted,
qualitatively similar to those obtained by Kusko {\it et al.} based on the 
$t$-$t'$-$t''$-$U$ model, see Fig.~2 in Ref.~[\onlinecite{Kusko}]. 
In order to realize the appearance of the pocket around
$({\pi\over 2},{\pi\over 2})$ at a relatively large doping, 
Kusko {\it et al.}\cite{Kusko} introduced the doping-dependent
$U$---$U_{eff}(\delta)$. Then the issue, whether the pocket around 
$({\pi\over 2},{\pi\over 2})$ really appears or not,
is dependent on the phenomenological form of $U_{eff}(\delta)$.
But the latter is not unambiguously clarified.
In the current $t$-$t'$-$t''$-$J$ model no parameters are introduced. 
Actually in our approach, the hoppings have been 
effectively multiplied by a factor $\delta$.
We notice that the pocket around 
$({\pi\over 2},{\pi\over 2})$ shows up naturally upon increasing doping.

We would also mention that, due to the intrinsic limitation of the MF treatment,
some ARPES data are not explained, for example, 
the development of the spectral weight inside the gap.
This needs the study beyond MF, as recently done by Kusunose and Rice.\cite{Kusunose}

So far we have studied the FS evolution with doping based on the $t$-$t'$-$t''$-$J$ 
model. Very recently the similar model was adopted by Li {\it et al.}\cite{Li} to 
calculate the spin dynamics in the SC and normal states around the optimal doping,
and consistent results with neutron scattering data on 
Nd$_{2-x}$Ce$_x$CuO$_4$ ($x=0.15$)\cite{Yamada} were obtained.
It shows that the $t$-$J$ type models may be available in the study of 
electron-doped cuprates. On the other hand, more theoretical predictions 
based on them are needed.
As an example, following Li {\it et al.} one may further calculate the spin 
dynamics in the AF state. As already shown above, within certain doping region the two 
energy bands are both crossed by the Fermi level. It is expected that the interband 
excitation will lead to characteristic spin susceptibility. Another interesting topic
is to investigate the potential spinon pairing under the AF background and the possible 
coexistence of the AF and SC states. Note that the pairing here is special 
due to the existence of two bands around the Fermi level. (Similar situation 
seems not to be present in the hole-doped case.) 
The studies on these issues are in progress.

In conclusion, the electron-doped $t$-$t'$-$t''$-$J$ model has been studied
in the AF state. By using slave-boson MF treatment we have calculated the two 
energy bands in the MBZ. It was shown that at low doping only one band is crossed 
by the Fermi level and small FS pockets appear 
around $(\pm\pi,0)$ and $(0,\pm\pi)$. Upon increasing doping, 
the other band will be crossed and new FS pockets form around 
$(\pm{\pi\over 2},\pm{\pi\over 2})$. The evolution of the FS with doping
and the calculated spectral weight are in good agreement with the ARPES data. 

\bigskip
This work was supported by the Texas Center for Superconductivity at the
University of Houston and the Robert A. Welch Foundation.

\end{document}